\documentclass[12pt]{article}
\usepackage{amsmath,amsthm,amscd,amsfonts,amssymb}
\usepackage{latexsym}
\usepackage[usenames]{color}

\newcommand{\NN}{\mathbb N}

\newcommand{\RR}{\mathbb R}
\newcommand{\TT}{\mathbb T}
\newcommand{\ZZ}{\mathbb Z}

\newcommand{\Znu}{\ZZ^\nu}

\newcommand{\nn}{\mathcal N}

\newtheorem{thm}{Theorem}[section]
\newtheorem{lemma}[thm]{Lemma}

\newtheorem{rem}[thm]{Remark}
\newtheorem{hyp}[thm]{Hypothesis}

\def\half{\frac{1}{2}}

\def\pf{{\noindent \bf Proof: }}
\begin{document}
\author{M Krishna \\ Institute of Mathematical Sciences \\ Taramani Chennai 600 113 \\ India \\
e-mail: krishna@imsc.res.in \\
Phone : (91)(44)22543309 \\
Fax   : (91)(44)22541586 }
\title{AC spectrum for a class of random operators at small disorder}
\date{}
\maketitle
\begin{abstract} 
In this paper we present a class of Anderson type operators 
with independent, non-stationary (non-decaying) random potentials
supported on a subset of positive density in 
the odd-dimensional lattice  and prove the existence of pure  
absolutely continuous spectrum in the middle of the band for
small disorder. 
\end{abstract}

\newpage

\section{Introduction}

Let the discrete Laplacian on $\ell^2(\Znu)$ 
be defined by
$$
(\Delta u)(n) = \sum_{i=1}^\nu (T_i + T_i^{-1})u(n)
$$
where 
$$
(T_i)u(n) = u(n-e_i), e_i \in \Znu, ~ e_{ik} = \delta_{ik}, ~ i,k=1,\dots,\nu. 
$$
Given some infinite subset $\nn \subset \Znu$ consider the random potentials
\begin{equation}\label{vomeg}
V^\omega = \sum_{k \in \nn} \omega_k \phi_k
\end{equation}
where $\{\phi_k, k\in \nn\}$ are real valued functions of 
compact and mutually disjoint supports  
and $\{\omega_k\}$ are independent identically distributed
random variables.  We then look at the model

\begin{equation}\label{themodel}
H_\lambda^\omega = \Delta + \lambda V^\omega
\end{equation}

and study its spectrum.

To specialize $V^\omega$ further we take the multiplication operators
\begin{equation}\label{quis}
(Q_iu)(n) = n_i u(n), i=1, \dots ,\nu  
\end{equation}
on $\ell^2(\Znu)$. We note that $T_i, T_i^{-1}$ are unitary for each $i$.
We define 
\begin{equation}\label{eqn1}
A = \half\sum_{i=1}^\nu \left\{Q_i\left(T_i^{-1} - T_i\right) + 
\left(T_i^{-1} - T_i\right)Q_i\right\}.
\end{equation}
The $Q_i$'s and $A$ are self-adjoint on $\ell^2(\Znu)$ with dense domains 
and the set of sequences in $\ell^2(\Znu)$
of finite support forms a core for all of them. 

Given a function $\phi$
we denote the operator of multiplication by $\phi$ on $\ell^2(\Znu)$
also by the same symbol and for the following we set
$$
\Lambda_s(n) = \{m \in \Znu : |m-n| \leq s \}.
$$

\begin{hyp}\label{hyp1}
We assume that there is an infinite subset $\nn \subset \Znu$ and
a collection of functions $\{\phi_k, ~ k \in \nn\}$ such that:
\begin{enumerate}
\item $\{\phi_k, k\in \nn\}$ are non-negative, 
bounded uniformly
by $1$, are of mutually disjoint supports and the quantities
$$
[A, \phi_k], ~ [A, [A, \phi_k]]
$$ 
are uniformly bounded in $k$. We set $\||[A, \phi_k]\||_\infty = 
\sup_{k \in \nn} \|[A, \phi_k]\|_\infty$.

\item For each $k \in \nn$ there is an $n \in \Znu$ such that
$$
\Lambda_{|n|^\half}(n) \subset \{n^\prime \in \Znu: \phi_k(n^\prime) = 1\}
$$ 

\item The distribution $\mu$ of the random variables 
$\{\omega_k: k \in \nn\}$ is  
compactly supported in $\RR$ and $0 \in \mathrm{supp}(\mu)$. 
We set 
$$
E_+ = \sup(\mathrm{supp}(\mu)), E_- = \inf(\mathrm{supp}(\mu)), ~
E_\infty = \mathrm{max}\{|E_+|, |E_-|\}
$$ 
\end{enumerate}
\end{hyp}

We then have the following theorem.

\begin{thm}\label{thm1}
Consider the random operators $H_\lambda^\omega$ given in equation 
(\ref{themodel}).  Suppose $V^\omega$ satisfies the hypothesis 
\ref{hyp1}.  Then
\begin{enumerate}
\item For each $\lambda \geq 0$, $\sigma_{ess}(H^\omega_\lambda ) = [-2\nu, 2\nu] + \lambda ~ \mathrm{supp} (\mu)$ a.e. $\omega$.
\item Let $\nu$ be odd and let $I$ be a closed interval contained in 
$(-2, 2)$.  Then
there is a $\lambda_I$ satisfying $\lambda_I E_\infty < 1$ such that for all 
$0 \leq \lambda < \lambda_I$, 
$$
\sigma_{s}(H_\lambda^\omega) \cap I = \emptyset, ~ a.e. \omega.
$$
\end{enumerate}
\end{thm}

\begin{rem}
\begin{itemize}
\item The hypothesis \ref{hyp1}(2) should not be necessary.  The boundedness
of the commutators stated in part (1) of the hypothesis should enable us
to prove that $\phi_k \geq \alpha > 0$ on a cube $\Lambda_{|n|^\half}(n)$
contained in the support of $\phi_k$ for large enough $k$ and this should
be sufficient to prove the first part of the theorem.  
\end{itemize}
\end{rem}

The spectral theory of random operators of the form given in 
equation (\ref{themodel}) is widely studied with various assumptions on 
$V^\omega$.  The spectrum is known to be pure point spectrum is well 
known for the Anderson model (which is the same as in 
equation (\ref{themodel}) when $\nn = \Znu, \phi_k(n) = \delta_{kn}$ )
when $\lambda$ is
large or at the edges of the spectrum. We refer to the book of 
Carmona-Lacroix \cite{carlac}, Cycone-Froese-Kirsch-Simon \cite{cfks},    
Figotin-Pastur \cite{fipa} and Stollmann \cite{sto} for the already
extensive literature on this aspect of the theory. 

There is a rich literature on the a.c. spectrum for decaying
random potentials on $\ell^2(\ZZ)$ with many sharp results.    
A review of some of these models is given by Denisov-Kiselev \cite{denkis}.

For the Anderson model, however, absolutely continuous spectrum 
is proving  to be elusive and the expected
result that there is such spectrum for small $\lambda$ and in higher
(than 2) dimensions is far from being realized.  The only
higher dimensional result for such a model is on the Bethe
lattice for which the absolutely continuous spectrum was shown by
Klein \cite{kl1} and by Froese-Hasler-Spitzer \cite{fhs}.  

In $\nu \geq 2$, a slightly modified model with decaying 
randomness (where in the
Anderson model one takes $a_n V^\omega(n)$ instead of $V^\omega(n)$
and requires $a_n \rightarrow 0, |n| \rightarrow \infty$ at some
rate) has been considered
by Krishna \cite{Kri1}, Anne Boutet de Monvel-Sahbani
\cite{as} and Bourgain \cite{bou}.  An alternative
collection of models are those for which $V^\omega(n)$ is zero
outside a "hyper surface" (of some thickness) in $\Znu$ as done
in Jaksi\'c - Last \cite{jl1}, \cite{jl2}. Yet another collection
of models assume that the $V^\omega(n)$ are zero outside a subset
$S$ of $\Znu$ which are "sparse" (i.e. of zero density in $\Znu$),
these are by Krishna \cite{Kri2}, Krutikov \cite{kru}, Molchanov \cite{mol}, 
Molchanov-Vainberg \cite{molv}. 

On the Bethe lattice Kupin \cite{kup} also considered decaying randomness
and showed a.c. spectrum. 

All these works show existence of absolutely continuous spectrum in some
region or the other of the spectrum.

\section{The proofs}

We give the proof of the theorem \ref{thm1} in this section.  We present the
ideas involved first.  

To show the statement (1) on the essential spectrum
we construct Weyl sequences for each point that is claimed to be in the 
essential spectrum.  We follow closely the ideas in Kirsch-Krishna-Obermeit \cite{kko}

The part (2) of the theorem is an application of Mourre theory
of the existence of of local conjugate, which is possible in the given
region of energy for small enough disorder parameter $\lambda$.  We 
need the dimension to be odd and also the energy to be small here
to show the positivity of commutators.

We start with a few technical lemmas first.

\begin{lemma}\label{lem1}
Let $\psi$ be a smooth function of compact support and let $H_\lambda^\omega$
be as in equation (\ref{themodel}) satisfying the hypothesis 
\ref{hyp1}(1),(2).  Then
$$
\|\psi(H_\lambda^\omega) - \psi(\Delta) \| \leq C |\lambda|
$$
where $C \leq E_\infty \int |t| |\widehat{\psi}(t)| ~ dt$.
\end{lemma}
\pf  Using the spectral theorem and the Fourier transform we have
$$
\psi(H_\lambda^\omega) - \psi(\Delta) = \int \left(e^{itH_\lambda^\omega} - e^{it\Delta}\right) ~ \widehat{\psi}(t) ~ dt.
$$
We also have by fundamental theorem of calculus
$$
e^{itH_\lambda^\omega} - e^{it\Delta} = \int_0^t e^{isH_\lambda^\omega}\lambda \sum_{k}\omega_k \phi_k~ e^{i(t-s)\Delta} ~ ds
$$
Putting these two equations together and estimating the operators norms, 
using the fact that
$$
\|\sum_{k} \omega_k \phi_k\|_\infty \leq \sup_{k} |\omega_k| \leq E_\infty, 
$$
from the hypothesis \ref{hyp1}(1),(2),  we get the lemma. \qed

For the following lemma we set 
$$
\Lambda(n) = \{m \in \Znu : |m-n| \leq |n|^\half \}.
$$


{\noindent \bf Proof of Theorem \ref{thm1} (1) : }
We essentially follow the ideas used in proving theorem 2.4 of 
Kirsch-Krishna-Obermeit \cite{kko} for doing this.  

Fix a $\lambda >0 $ and an $r \in \mathrm{supp}(\mu)$ and 
$E \in (-2\nu, 2\nu)$, we will show 
$$
E+\lambda r \in \sigma(H_\lambda^\omega), ~ a.e. ~ \omega.
$$
Given $\ell \in \NN$, we have
$$
\mu((r-\frac{1}{\ell}, r+\frac{1}{\ell})) > 0,
$$
from the definition of support of $\mu$.
Now consider the events 
$$
A_{k,\ell} = \{\omega: \omega_k \in (r-\frac{1}{\ell}, r+\frac{1}{\ell}) \}.
$$   
All these (mutually independent) events have (the same) positive 
probability for each fixed $\ell$ as $k \in \nn$ 
varies. 
Hence we have, for each fixed $\ell$,
$$
\sum_{k \in \nn} \mathrm{Prob}(A_{k, \ell}) = \infty,
$$
therefore by Borel-Cantelli lemma the events 
$\{A_{k,\ell}\}$ occur infinitely often with probability
one.  That is the set 
$$
\Omega_{\ell} = \bigcap_{r=1}^\infty \bigcup_{|k| \leq r} A_{k, \ell}
$$
has measure $1$.  Therefore the set
$$
\Omega_0 = \bigcap_{\ell \in \NN} \Omega_{\ell}
$$
also has measure $1$, being a countable intersection of measure 1 sets.  

Now since $E \in (-2\nu, 2\nu)$ which is the essential spectrum of $\Delta$,
there is a sequence (as seen for example using density of compactly 
supported functions in $\ell^2(\Znu)$ together with Theorem 7.2, 
Weidman \cite{wei}) $f_j$ of 
compactly supported functions in $\ell^2(\Znu)$,
with $\|f_j\|=1$, such that
$$
\|(\Delta - E) f_j\| \rightarrow 0, ~ \mathrm{as} ~ j \rightarrow \infty.
$$
Since $\Delta$ commutes with translations, it is also true that
for any $m \in \Znu$ the translates $f_j(\cdot - m)$ also satisfy the
above condition.  Given $\epsilon >0$ ,we find an $\ell$ such that
$\frac{1}{\ell} < \epsilon$, and a $j(\ell)$ such that
$$
\|(\Delta - E) f_{j(\ell)}(\cdot-m)\| \leq \epsilon, 
$$ 
with $f_{j(\ell)}$ having compact support and the size of this support
being the same for all $f_{j(\ell)}(\cdot -m)$ as $m$ varies. 
Now let $\omega \in \Omega_0$ be arbitrary but fixed, then the set 
$$
\nn_\omega = \{ k \in \nn : \omega_k \in (r - \frac{1}{\ell}, r + \frac{1}{\ell})\},
$$
is of infinite cardinality. The supports of $\phi_k$ are disjoint
by hypothesis \ref{hyp1}(1) so by hypothesis \ref{hyp1}(2) there is 
an $m(k) \in \Znu$ such that the sets 
$$
\Lambda_{m(k)^\half}(m(k)), ~ k \in \nn_\omega
$$
are mutually disjoint implying that the size of these sets
goes to infinity as $k$ goes to infinity in $\nn_\omega$. (Reason:
$|m(k)| \rightarrow \infty$ as $|k| \rightarrow \infty$).
Hence given $f_{j(\ell)}$ with compact support, we can find  
a $k_\ell \in \nn_\omega$ and an associated $m(k_\ell) \in \Znu$ 
such that
$$
\mathrm{supp}(f_{j(\ell)}(\cdot - m(k_\ell)) \subset \Lambda_{m(k_\ell)^\half}(m(k_\ell)), ~ \phi_{k_\ell} = 1 ~ \mathrm{on} ~ \mathrm{supp} (f_{j(\ell)}(\cdot - m(k_\ell))).
$$
Therefore we have for this $m(k_\ell) \in \Znu$,
\begin{eqnarray*}
\|H_\lambda^\omega - (E+\lambda r) )f_{j(\ell)}(\cdot -m(k_\ell))\|
 & \leq & \|(\Delta - E)f_{j(\ell)}(\cdot -m(k_\ell))\| \\ 
& ~ + & \|(\lambda \omega_{k_\ell}\phi_{k_\ell} - \lambda r) f_{j(\ell)}(\cdot -m(k_\ell)) \|
\\ & ~ \leq &  \epsilon +\frac{\lambda}{\ell} \leq (1+\lambda)\epsilon.
\end{eqnarray*}
This exhibits a Weyl
sequence $g_{\ell} = f_{j(\ell)}(\cdot -m(k_\ell))$ associated with the operators 
$H_\lambda^\omega$ for the point
$E + \lambda r$ showing that this point is in the essential spectrum of 
$H_\lambda^\omega$.  This proves the theorem for each $\omega \in \Omega_0$ 
as we vary $E \in [-2\nu, 2\nu]$ and $r \in \mathrm{supp}(\mu)$.  \qed

\vspace{.5cm}

{\noindent \bf Proof of Theorem \ref{thm1} (2) : }
We use Mourre theory for proving this.  We show that the operator $A$
defined in equation (\ref{eqn1}) is a local conjugate for $H_\lambda^\omega$
for all $\omega$ and $0 \leq \lambda < \lambda_I$.

We first verify Mourre's conditions (1) - (4) given in 
Definition 3.5.5 of \cite{demkri},
to see that the operator $A$ is a local conjugate of $H_\lambda^\omega$.

The conditions (1) - (3) of Definition 3.5.5 in \cite{demkri} 
are easy in view of the fact that $H_\lambda^\omega$
is a bounded operator, $[A, H_\lambda^\omega] = [A, \Delta] + [A, \lambda V^\omega]$ and 
$[A,[A, H_\lambda^\omega]] = [A,[A, \Delta]] + [A,[A, \lambda V^\omega]]$ are bounded by a simple computation and by hypothesis \ref{hyp1}(1)
for any $\lambda \geq 0$ and any $\omega$.  

Therefore we are left only to verify the Mourre estimate 
(bound in (4) of Definition 3.5.5 in \cite{demkri})
 to conclude that $\sigma_{sc}(H_\lambda^\omega) = \emptyset
$ for these $\lambda, \omega$ from the theorem of Mourre (see theorem 3.5.6, \cite{demkri})). 

Let $P_{H_\lambda^\omega}(I)$ denote the spectral 
projection of $H_\lambda^\omega$ associated with the interval 
$I = [a, b] \subset (-2,2)$. Let $\delta = \min\{1+a/2, 1-b/2\}$. 
Since $0 \in \mathrm{supp}(\mu)$ by
the hypothesis \ref{hyp1}(2),
part (1) of the theorem ensures that $(-2, 2) \subset
\sigma_{ess}(H_\lambda^\omega)$, so this spectral projection is non-trivial.
We will show that the bound 
\begin{equation}\label{eqn2}
P_{H_\lambda^\omega}(I) [A, H_\lambda^\omega] P_{H_\lambda^\omega}(I) \geq
3\delta P_{H_\lambda^\omega}(I) 
\end{equation}
is valid for all $0 \leq \lambda < \lambda_I$ for some $\lambda_I$.

We consider a smooth
function $\psi$, which is identically $1$ on $I =[a,b]$  
and zero outside $(-1+a/2, 1+b/2)$. 
Then it is clear that $\psi = 0$ outside $(-2 + \delta, 2- \delta)$
for the $\delta$ given above.
We then have as $\lambda \rightarrow 0$, 
\begin{equation}\label{eqn3}
\begin{split}
\psi(H_\lambda^\omega) [A, H_\lambda^\omega] \psi(H_\lambda^\omega) 
&= \psi(\Delta) [A, \Delta] \psi(\Delta) + (\psi(H_\lambda^\omega)-  \psi(\Delta))[A, \Delta] \psi(H_\lambda^\omega) \\ 
& + ~ ~ \psi(\Delta)[A, \Delta](\psi(H_\lambda^\omega)-  \psi(\Delta)) \\ 
& +  ~  ~ \psi(H_\lambda^\omega) \lambda \sum_{k} \omega_k [A, \phi_k] \psi(H_\lambda^\omega) \\
= \psi(\Delta) [A, \Delta] \psi(\Delta) + O(|\lambda|)
\end{split}
\end{equation}
where the last statement follows from Lemma \ref{lem1}, the uniform boundedness
of 
$$
\|\sum_{k} \omega_k [A, \phi_k]\| \leq E_\infty \||[A, \phi_k]\||_\infty
$$
coming from hypothesis \ref{hyp1}(1),(2). 

Computing the commutator $[A, \Delta]$ we get 
$$
\psi(\Delta) [A, \Delta] \psi(\Delta) = - \psi(\Delta)^2 \sum_{j}(T_j - T_j^{-1})^2.
$$
Using the Fourier transform we see that the above operator is unitarily equivalent to the operator of multiplication by the function
$$
\left(\psi(2\sum_{i=1}^\nu \cos(\theta_i))\right)^2 4\sum_{i=1}^\nu \sin^2 (\theta_i)
$$
on $L^2(\TT^\nu, d\sigma)$, where $\sigma$ is the normalized Lebesgue measure
on the $\nu$ dimensional torus $\RR^\nu/\ZZ^\nu$.  

To estimate this quantity from below,  let 
$$
W = \{\theta \in \TT^\nu : \psi(2\sum_{i=1}^\nu \cos(\theta_i)) \neq 0\}
$$
Since $\psi$ is zero outside $(-2 + \delta, 2-\delta)$, we see that
$$
W \subset 
\{\theta \in \TT^\nu : -1 + \delta/2 <\sum_{i}\cos(\theta_i) < 1 - \delta/2\}.
$$
Then we claim that for every $\theta \in W$, there is an index
$j$ such that 
$$
-1 + \delta/2 \leq \cos\theta_j \leq 1 - \delta/2. 
$$
Suppose this is not the case and there is a $\theta^0 \in W$ such that
$$
\cos(\theta^0_i) > 1 - \delta/2 ~ \mathrm{or} ~ \cos(\theta^0_i) < -1 + \delta/2, ~ \mathrm{for ~ all} ~ i = 1, \dots, \nu.
$$
Let 
$$
K_\pm = \{ i : \pm \cos(\theta^0_i) > 0\} ~ \mathrm{and} ~ n_{\pm} = \# K_\pm.
$$
Then either $n_+ > n_-$ or $n_- > n_+$ since the dimension $\nu$ is odd.
Consider the case $n_+ > n_-$ (the argument for the other case is similar).
Then we have
$$
\sum_{i} \cos(\theta^0_i) = \sum_{i \in K_+} \cos(\theta^0_i) 
+ \sum_{i \in K_-} \cos(\theta^0_i) > n_+(1-\delta/2) - n_- \geq 1 - \delta/2 
$$
contradicting the fact that $\theta^0 \in W$.

Therefore we have for every $\theta \in W$,for some $j_\theta$,
$$
4\sum_{i=1}^\nu \sin^2(\theta_i) \geq 4\sin^2(\theta_{j_\theta}) =4( 1 - \cos^2(\theta_{j_\theta})) \geq 4(1 - (1 - \delta/2)^2)  \geq 3 \delta,
$$
since $\delta < 1$.
Thus we have as $\lambda$ goes to zero, again using Lemma \ref{lem1},
$$
\psi(\Delta) [A, \Delta] \psi(\Delta) \geq 3\delta \psi(\Delta)^2
= 3\delta \psi(H_\lambda^\omega) + O(\lambda).
$$
Putting this inequality together with the inequality (\ref{eqn3})
we get that for sufficiently small $\lambda$,
$$
\psi(H_\lambda^\omega) [A, H_\lambda^\omega] \psi(H_\lambda^\omega) \geq 
3 \delta \psi(H_\lambda^\omega)^2 + O(\lambda).
$$
Now we multiply either side on the inequality by $P_{H_\lambda^\omega}(I)$ 
and choose $\lambda_\delta$ such that the terms $O(\lambda)$ is smaller than
$\delta$ for each $\lambda < \lambda_\delta$, to get 
$$
P_{H_\lambda^\omega}(I) [A, H_\lambda^\omega])P_{H_\lambda^\omega}(I) \geq
2 \delta P_{H_\lambda^\omega}(I),  
$$
obtaining the inequality (\ref{eqn2}) with $\lambda_I = \lambda_\delta$.

The absence of point spectrum in $I$ follows from the Virial theorem
(see a version given in Proposition 2.1, \cite{Kri1},  the condition
(1) there is easy since $[A, H_\lambda^\omega]$ extends to a bounded
operator from $D(A)$ and (2) there holds for all vectors in $\ell^2(\Znu)$ of 
finite support, which are mapped into domain of $A$ by 
$(H_\lambda^\omega \pm z)^{-1}$ for $|z| > \|H_\lambda^\omega\|$ as
can be seen from a use of Neumann series expansion).
\qed
 
\section{Examples}

We give examples of random potentials that satisfy the hypothesis \ref{hyp1},
our examples are adapted from the continuum case given in Krishna \cite{Kri1}
which are extensions of the Rodnianski-Schlag models \cite{rs}.
We note that we use below the $\ell^2$-norm on $\RR^\nu$ while we use the
$\ell^\infty$ norm on $\Znu$, so that
$$
|x| = \sqrt{\sum_{i=1}^\nu x_i^2}, ~ x \in \RR^\nu ~ \mathrm{and} ~ 
|n| = \max\{ |n_i|, i=1, \dots, \nu\}, ~ n \in \Znu.
$$
Recall the definition  
$$
\Lambda_r(n) = \{ m \in \Znu : |m-n| \leq r \}.
$$
We define a function $\phi_{r,n}$ associated with a set $\Lambda_r(n)$
 on $\Znu$ as follows.  Let $\phi$ be a smooth bump function on $\RR^\nu$
with (full) support in $\{x : |x| < 1\}$ vanishing on the sphere
$\{x : |x| = 1\}$.  Let
$$
\widetilde{\phi_{r,n}}(x) =  \phi(\frac{x-n}{r}), ~ x\in \RR^\nu
$$
and let $\phi_{r,n}$ be its restriction to $\Znu$.  For
any $n \in \Znu$, let $r(n)$ be a positive number satisfying
$$
c_1 |n| \leq r(n) \leq c_2 |n|, ~ 0 < c_1 \leq c_2 < \infty,
$$
for any $n \in \Znu$.  Then for any $n$ we claim that 
$[A, \phi_{r(n), n}]$ and 
$[A,[A, \phi_{r(n),n}]]$ are uniformly bounded in $n$.
We will show the boundedness of the second commutator and the computation
for the first commutator becomes clear in the process.  We will again show the
boundedness of 
$$ 
 \left[ \sum_{j=1}^\nu Q_jT_j , \left[ \sum_{k=1}^\nu Q_kT_k, \phi_{r(n), n}\right] \right], 
$$
the other terms in the expansion of $[A, [A, \phi_{r(n), n}]]$ are 
similarly computed and shown to be bounded.  
\begin{equation}
\begin{split}
 & \left[ \sum_{j=1}^\nu Q_jT_j , \left[ \sum_{k=1}^\nu Q_kT_k, \phi_{r(n), n}\right] \right] \\ & = \sum_{j,k=1}^\nu Q_j \left[T_j, Q_k [T_k, \phi_{r(n), n}]\right]\\
& =  \sum_{j,k=1}^\nu Q_j [T_j, Q_k] [T_k, \phi_{r(n), n}] + Q_j Q_k [T_j, [T_k, \phi_{r(n), n}]]\\
& =  \sum_{j,k=1}^\nu Q_j \delta_{jk} [T_k, \phi_{r(n), n}] + Q_j Q_k [T_j, [T_k, \phi_{r(n), n}]]\\
& =  \sum_{k=1}^\nu Q_k [T_k, \phi_{r(n), n}] + \sum_{k,j=1}^\nu Q_j Q_k [T_j, [T_k, \phi_{r(n), n}]]\\
\end{split}
\end{equation}

We note that the vectors $e_k \in \Znu \subset \RR^\nu, k=1,\dots, \nu$ 
are also unit vectors along the co-ordinate axes and we use this fact 
without further comment in the calculations below.

Let $u \in \ell^2(\Znu)$, then
\begin{equation}\begin{split}
([T_k, \phi_{r(n), n}]u)(x) &= \left(\phi_{r(n), n}(x-e_k) - \phi_{r(n), n}(x) \right)(T_k u)(x) \\
&= \left(\phi(\frac{x-n -e_k}{r(n)}) - \phi(\frac{x-n}{r(n)})\right)(T_k u)(x) \\
&= \left(\frac{-1}{r(n)} \frac{\partial}{\partial t} \phi(\frac{x-n}{r(n)} - t_0 e_k) \right) (T_k u)(x) 
\end{split}
\end{equation}
for some $t_0 \in (0, \frac{1}{r(n)})$, using the mean value theorem for the function $\phi$ on the 
line segment in the direction of $\{ \frac{x-n}{r(n)} - t e_k, 0 \leq t \leq \frac{1}{r(n)} \}$ in $\RR^\nu$. From this we see that 
\begin{equation}\begin{split}\label{eqn9}
|Q_k ([T_k, \phi_{r(n), n}]u)(x)| &= |\frac{x_k - e_k}{r(n)}\left(\frac{1}{r(n)} \frac{\partial}{\partial t} \phi(\frac{x-n}{r(n)} - t_0 e_k) \right) (T_k u)(x)| \\ 
&\leq C (T_ku)(x)| ,
\end{split}\end{equation} 
the bound coming from the uniform bound on the partial derivative of $\phi$ in
the unit ball and from the uniform boundedness of $(x-n+e_k)/r(n)$ coming 
from the assumption on $r(n)$.  Similarly we find using the mean value theorem
\begin{equation}\begin{split}
& ([T_j,[T_k, \phi_{r(n), n}]]u)(x) \\ & = \left(\left(\phi_{r(n),n}(x-e_k-e_j) - \phi_{r(n),n}(x-e_j)\right) \right . \\ & - \left . \left(\phi_{r(n),n}(x-e_k) - \phi_{r(n),n}(x)\right)\right) (T_jT_ku)(x)
\\
& = \left(\frac{-1}{r(n)} (\frac{\partial}{\partial t} \phi)(\frac{x-n-e_j}{r(n)}-t_0e_k) +
 \frac{1}{r(n)} (\frac{\partial}{\partial t} \phi)(\frac{x-n}{r(n)}-t_1e_k) \right)
 (T_jT_ku)(x) \\
& = \left(\left(\frac{-1}{r(n)} (\frac{\partial}{\partial t} \phi)(\frac{x-n}{r(n)}-\frac{1}{r(n)}e_j-t_0e_k) \right.
+ \frac{1}{r(n)} (\frac{\partial}{\partial t} \phi)(\frac{x-n}{r(n)}-t_0e_k)\right) \\
& + \left(\frac{-1}{r(n)} (\frac{\partial}{\partial t} \phi)(\frac{x-n+}{r(n)}-t_0e_k) +\right. 
\left.\left. \frac{1}{r(n)} (\frac{\partial}{\partial t} \phi)(\frac{x-n}{r(n)}-t_1e_k) \right)\right)
 (T_jT_ku)(x),
\end{split}\end{equation} 
for some $t_0, t_1 \in (0, \frac{1}{r(n)})$.  Applying the mean value theorem
one more time we get 
\begin{equation}\begin{split}
& ([T_j,[T_k, \phi_{r(n), n}]]u)(x) \\  &= 
\left(\frac{1}{r(n)^2} (\frac{\partial^2\phi}{\partial t \partial s}\right.
(\frac{x-n}{r(n)}-s_1e_j - t_0e_k)) \\ 
& + \frac{(t_0-t_1)}{r(n)} (\frac{\partial^2 \phi}{\partial t^2})
\left. (\frac{x-n}{r(n)}- t_2e_k) \right) (T_jT_ku)(x),
\end{split}\end{equation} 
where $t_2 \in (t_1, t_0)$ (taking w.l.g. $t_1 < t_0$ ) with the notation
that the variable $s,t$ denotes taking derivatives along the 
directions $e_k, e_j$ respectively.  Denoting below these partial derivatives
by $\frac{\partial}{\partial x_j}$ and  $\frac{\partial}{\partial x_k}$ 
respectively we get the bound
$$
\|([T_j,[T_k, \phi_{r(n), n}]]u)\| \leq \frac{2}{r(n)^2} 
(\|\frac{\partial}{\partial x_j}\frac{\partial}{\partial x_k}(\phi)\|_\infty + 
\|\frac{\partial^2}{\partial x_k^2}(\phi)\|_\infty ), 
$$ 
where we used the fact that $|t_1 - t_0|\leq 2/r(n)$.  From this bound
and noting that the coordinates $x_j, x_k$ are bounded
in modulus by $c r(n)$ when $x = (x_1, \dots, x_\nu)$ is in the
support of $\phi_{r(n),n}$ and its derivatives. Therefore the quantity 
$$
(Q_j Q_k [T_j, [T_k, \phi_{r(n),n}]])u(x) = x_j x_k [T_j, [T_k, \phi_{r(n),n}]])u(x)
$$  
has the bound 
$$
\|(Q_j Q_k [T_j, [T_k, \phi_{r(n),n}]])u\| \leq c^2  r(n)^2 \frac{2}{r(n)^2}
(\|\frac{\partial}{\partial x_j}\frac{\partial}{\partial x_k}(\phi)\|_\infty +
\|\frac{\partial^2}{\partial x_k^2}(\phi)\|_\infty )\|u\|.
$$
This gives the boundedness of $[A, [A, \phi_{r(n),n}]]$ uniformly in 
$n$.

Using these facts we construct a random potential as
follows.  Let $M$ be a large integer and let $A_0 =\{m \in \Znu : |m| \leq M\}$
and consider the annuli 
$$
A_{k+1} = \{ m \in \Znu : 2^k M < |m| \leq 2^{k+1}M \}, ~ k=0,1,\dots.
$$ 
We define  
$$
r(m) = 2^{k-2} M, ~ \forall m \in A_k, ~ k=1,2,\dots,
$$ 
to get the uniform bounds
$$
4 \leq \frac{2^k M}{2^{k-2}M} \leq \frac{|n|}{r(n)} \leq \frac{2^{k+1} M}{2^{k-2}M} \leq 8, ~ \mathrm{for ~ all } ~ n \in A_k, ~k=1,2,\dots.
$$
Then consider the sets $\nn_k, ~k=1,2,\dots,$ such that 
$$
\nn_k \subset A_{k} , ~ n,m \in \nn_k \implies 
\Lambda_{r(n)}(n) \cap \Lambda_{r(m)}(m) = \emptyset. 
$$
It is clear that the cardinality of $\nn_k$ is at least $2\nu$, since one can 
fit a cube $\Lambda_{r(n)}(n)$ alongside of each face of the cube
$\Lambda_{2^kM}(0)$ in $A_{k+1}$.

Let $\nn = \cup_{k=1}^\infty \nn_k$ and let $\{\omega_\ell : \ell \in \nn\}$
be i.i.d. random variables with distribution $\mu$.
Then by the definition of $A_k$ and $r(n)$ it is
clear that $c_1 |n| \leq r(n) \leq c_2 |n|$ is valid.  The random
potential  
$$
V^\omega = \sum_{n \in \nn} \omega_n \phi_{r(n), n}
$$ 
then satisfies the hypothesis (\ref{hyp1}). 

The distribution $\mu$ can be atomic, singularly continuous or absolutely
continuous or a mixture of all these.


\begin{thebibliography}{ll}

\bibitem{bou} J. Bourgain
\textit{Random Lattice Schr\"odinger Operators with Decaying Potential: Some
Higher Dimensional Phenomena} V D Milman and G Schechtman (Eds), Lecture Notes in Mathematics 1807, 70-98 (2003).

\bibitem{as}A. Boutet de Monvel and  J. Sahbani: 
\textit{ On the spectral properties of discrete Schr\"odinger operators: the multi-dimensional case}, Rev. Math. Phys. {\bf 11},  1061--1078(1999)

\bibitem{carlac}
R. Carmona and J. Lacroix 
  \textit{Spectral Theory of Random Schr\"odinger Operators,}
  (Birkh\"auser Verlag, Boston 1990)

\bibitem{denkis} S. A. Denisov and A. Kiselev 
\textit{Spectral properties of Schrodinger operators with decaying potentials},
Spectral Theory and Mathematical Physics:
A Festschrift in Honor of Barry Simon's 60th Birthday,
Proceedings of Symposia in Pure Mathematics, Volume: 76
(AMS, Providence, 2007)

\bibitem{demkri}
M. Demuth and M. Krishna 
\textit{Determining spectra in Quantum Theory}, Progress in Mathematical Physics Vol 44, (Birkhauser, Boston, 2005).

\bibitem{cfks} H. Cycon, R. Froese, W. Kirsch, B. Simon 
  \textit {Schr\"odinger Operators}, 
  {Texts and Monographs in Physic} (Springer Verlag, 1985)

\bibitem{fipa} A. Figotin and L. Pastur: \textit {Spectra of Random and Almost-Peri
odic Operators} (Springer Verlag, Berlin 1992)

\bibitem{frsa} R. L. Frank and A. O. Safronov
\textit{Absolutely continuous spectrum of a class of random non ergodic
   Schr\"odinger operators}
Int. Math. Res. Not. vol  {\bf 42},  2559--2577 (2005)


\bibitem{fhs} R. Froese, D. Hasler, W. Spitzer 
\textit{Absolutely continuous spectrum for the Anderson model on a tree: 
a geometric proof of Klein's theorem. }
Rev. Math. Phys. {\bf 21}, No. 6, 709--733 (2009)


\bibitem{fs} J. Fr\"ohlich and T. Spencer 
\textit{Absence of diffusion in the Anderson tight binding model for large disorder
 or low energy}, Comm. Math. Phys. {\bf 88}, 151--184 (1983)

\bibitem{jl1}V. Jaksi\'c and Y. Last 
\textit{ Corrugated surfaces and a.c. spectrum
}, Rev. Math. Phys. {\bf 12},  1465--1503(2000)

\bibitem{jl2}V. Jaksi\'c and Y. Last 
\textit{ Surface states and spectra}, Comm. Math. Phys. {\bf 218},  459--477 (2001)

\bibitem{kko}
W. Kirsch, M. Krishna, J. Obermeit 
    \textit{Anderson model with decaying randomness-- mobility edge,} Math. Zeitschrift, {\bf 235}, 421--433(2000)

\bibitem{kl1} A. Klein 
       \textit{Extended states in the Anderson model on the Bethe lattice,}
        Adv. Math. {\bf 133}, 163--184 (1998)

\bibitem{Kri1}
M Krishna
\textit{Absolutely continuous spectrum and spectral transition for some continuum random operators}, to appear in Proc. Ind. Acad. Sci. (2011)

\bibitem{Kri2}M. Krishna  
\textit{ Absolutely continuous spectrum for sparse potentials}
, Proc. Ind. Acad. Sci {\bf 103},  333-339.

\bibitem{kru}
D. Krutikov 
\textit{Schr\"odinger operators with random sparse potentials, existence of wave
operators}, Lett. Math. Phys. {\bf 67}, 133-139(2004)

\bibitem{kup} S. Kupin
\textit{Absolutely continuous spectrum of a Schr\"odinger operator on a tree}, J. Math. Phys. {\bf 49}, 113506.1-113506.10 (2008)

\bibitem{mol}S. Molchanov 
\textit{ Multiscattering on sparse bumps}, in \textit { Advances in Differential Equations and Mathematical Physics (Atlanta, GA, 1997)}, 157--181, Contemp. Math., 217, Amer. Math. Soc., Providence, RI, 1998

\bibitem{molv}S. Molchanov\ and\ B. Vainberg 
\textit{ Spectrum of multidimensional Schr\"odinger operators with sparse potentials}, in \textit { Analytical and Computational Methods in Scattering and Applied Mathematics (Newark, DE, 1998)}, 231--254, Chapman \& Hall/CRC, Boca Raton, FL, 2000

\bibitem{rs} I. Rodnianski and W. Schlag
\textit{Classical and quantum scattering for a class of long range random potentials.}  Int. Math. Res. Not.  {\bf 5},243–300 (2003)

\bibitem{sto} P. Stollmann 
    \textit {Caught by Disorder, Bound states in random media}, {Progress in Mat
hematical Physics}  20 (Birkh\"auser Verlag, Boston MA 2001)

\bibitem{wei} J. Weidman
\textit{Linear Operators in Hilbert Spaces} GTM-68 (Springer Verlag, Berlin, 1987)
\end{thebibliography}
\end{document}